

\magnification=\magstep2
\baselineskip=12pt
\overfullrule=0pt
\hsize = 4.5 in
\vsize = 6.5 in
\language=0
\centerline{\bf Equations of Associativity}
\centerline{\bf in Two-Dimensional Topological
Field Theory  }
\centerline{\bf as Integrable Hamiltonian Nondiagonalizable }
\centerline{\bf Systems of Hydrodynamic Type \footnote \dag{\rm
This work was partially supported by the International Science
Foundation (Grant No. RKR000) and the Russian Foundation of
Fundamental Researches (Grant No. 94--01--01478 --- O.I.M. and
Grant No. 93--011--168 --- E.V.F.)} }
\vskip .4 in
\centerline{\bf O.I.Mokhov}
\vskip .1 in
\centerline{\it Department of Geometry and Topology}
\centerline{\it Steklov Mathematical Institute}
\centerline{\it ul. Vavilova, 42}
\centerline{\it Moscow, GSP-1, 117966, Russia}
\centerline{\it e-mail: mokhov@mian.su}

\vskip .1 in
\centerline{\bf and}
\vskip .1 in
\centerline{\bf E.V.Ferapontov }
\vskip .1 in
\centerline{\it Institute of Mathematical Modelling}
\centerline{\it Miusskaya, 4}
\centerline{\it Moscow, 125047, Russia}
\centerline{\it e-mail: fer\#9@imamod.msk.su}
\vskip .6 in
\def \grad {\rm grad\ }
\def \diag {\rm diag\ }
\def \here {\it here\ }
\centerline{\bf \S \ 1.  Introduction}

\vskip .2 in

Let us consider a function of $n$ independent variables
$F(t^1,...,t^n)$ satisfying the following two
conditions:
\item{1.} {\it The matrix $$\eta _{\alpha \beta} =
{\partial ^3 F \over \partial t^1 \partial t^{\alpha} \partial t^{\beta}}
\ \ \ \ \ \ (\alpha, \beta = 1,...,n)$$
is constant and nondegenerate.}

Note that the matrix $\eta _{\alpha \beta}$ completely
determines dependence of the function $F$ on
the fixed variable $t^1$.

\item{2.} {\it For all $t=(t^1,...,t^n)$ the functions
$$c^{\alpha}_{\beta \gamma} (t) =
\eta ^{\alpha \mu} {\partial ^3 F \over \partial t^{\mu}
\partial t^{\beta} \partial t^{\gamma}} \ \ \ \ \ \
(\here \ \  \eta ^{\alpha \mu} \eta _{\mu \beta} =
\delta ^{\alpha}_{\beta})$$
are structural constants of an
associative algebra
 $A(t)$ in $n$-dimensional space with a basis
$e_1,...,e_n$ and the multiplication
$$ e_{\beta} \circ e_{\gamma} = c^{\alpha}_{\beta \gamma} (t) e_{\alpha}.$$}


The conditions 1 and 2 impose a complicated
overdetermined system of nonlinear partial differential equations of
the third order on the function $F$. This system
is known in two-dimensional topological field
theory as the equations of associativity or the
Witten-Dijkgraaf-H.Verlinde-E.Verlinde (WDVV) system ([10--13],
all necessary physical motivations and theory of integrability
for the equations of associativity can be found in the survey of
B.Dubrovin [1]).

For $n=3$ two essentially different types of dependence of
the function $F$ on the fixed variable $t^1$ were considered
by Dubrovin:

$$F= {1 \over 2} (t^1)^2 t^3 + {1 \over 2} t^1 (t^2)^2 + f(t^2,t^3)$$

and

$$F={1 \over 6} (t^1)^3 + t^1 t^2 t^3 + f(t^2,t^3).$$

For these cases the equations of associativity reduce to the
following two nonlinear equations of the third order for a function
$f=f(x,t)$ of two independent variables:

$$f_{ttt} = f^2_{xxt} - f_{xxx} f_{xtt} \eqno{(1.1)}$$

and

$$f_{xxx} f_{ttt} - f_{xxt} f_{xtt} = 1  \eqno{(1.2)}$$

correspondingly.

Following  [2,3], introduce new variables
$$a = f_{xxx}, \ b = f_{xxt}, \ c =f_{xtt}.$$

As it was shown in the papers [2,3], in new variables the equations (1.1) and
(1.2)
has the form of
 $3 \times 3 \ $
systems of hydrodynamic type:

$$\cases {a_t = b_x,&\ \cr
b_t = c_x,&\ \cr
c_t = (b^2 - ac)_x &\ \cr} \eqno{(1.3)}$$

and

$$\cases {a_t =b_x,&\ \cr
b_t = c_x,&\ \cr
c_t = \bigl ( {{(1+bc)} / a} \bigr )_x &\ \cr} \eqno{(1.4)}$$

correspondingly.

We recall that systems of hydrodynamic type are by definition
systems of quasilinear equations of the form
$$u^i_t = v^i_j (u) u^j_x.$$

The main advantage of representation of the equations of
associativity in the form (1.3), (1.4) is the presence of
efficient and elaborate theory of integrability of systems of
hydrodynamic type --- see, for example, the surveys of Tsarev [4],
Dubrovin and Novikov [5], and also the papers [6--9] devoted to
systems of hydrodynamic type, which do not possess Riemann invariants.

Everywhere in this paper we consider only strictly hyperbolic systems,
i.e., the eigenvalues of the matrix $v^i_j$ are real and distinct.
Thus, both the systems under consideration (1.3) and (1.4) are strictly
hyperbolic.

In   \S \  2 a Hamiltonian property of the system (1.3) is established.
For this system a local nondegenerate Hamiltonian structure
of hydrodynamic type (a Poisson bracket of Dubrovin-Novikov type [5])
is found. In contrast to (1.3), the integrable system of hydrodynamic type
(1.4) possesses only nonlocal Hamiltonian structures of hydrodynamic type
(see [14--16]).
Investigation, which was made in [2,3], showed
that both these systems (1.3) and (1.4) are nondiagonalizable
(i.e., do not possess Riemann invariants).

In  \S \ 3 we consider general theory of integrability
of nondiagonalizable Hamiltonian $3 \times 3$ systems of
hydrodynamic type following [6-9].
It turns out, that any such a system can be reduced to
the integrable three wave system by some standard
chain of transformations. We shall demonstrate this procedure
for the system (1.3). Correspondingly, it is shown that any
solution of the integrable three wave system generates solutions
of the equation of associativity (1.1).

In  \S \ 4 an explicit B\"acklund type transformation connecting
solutions of the systems (1.3) and (1.4) is found.

\vskip .3 in

\centerline{\bf \S \ 2.  Hamiltonian representation of the system (1.3)}

\vskip .2 in

As it was noticed by Dubrovin [1], the equation (1.1)
is connected with a spectral problem, which
has the following form in the variables $a, b, c$:

$$\Psi _x = z A \Psi = z {\pmatrix
{0&1&0 \cr
b & a & 1 \cr
c & b & 0 \cr}} \Psi, $$

$$\Psi _t = z B \Psi = z {\pmatrix
{0&0&1 \cr
c & b & 0 \cr
b^2 -ac & c & 0 \cr}} \Psi.  \eqno{(2.1)}$$

Compatibility conditions for the spectral problem (2.1)
are equivalent to the following two relations between the matrices
 $A$ и $B$:

$$\cases {A_t = B_x, &\ \cr
[A,B]=0, &\ \cr}   \eqno{(2.2)}$$
which are satisfied identically by virtue of the equations (1.3)
(here [\ \ ,\ \ ] denotes the commutator).
\vskip .1 in

{\bf Lemma 1}. {\it The eigenvalues of the matrix $A$
are densities of conservation laws of the system (1.3).}

\vskip .1 in

{\bf Proof}. So the matrices $A$ and $B$ commute and have simple spectrum,
they can be diagonalized simultaneously
$$A= PUP^{-1},\ \ \ B=PVP^{-1}.$$
Here $U=\diag (u^1,u^2,u^3),\ \ V=\diag (v^1,v^2,v^3).$
Substitution in the equation (2.2) gives
$$[P^{-1} P_t, U] + U_t = [P^{-1}P_x, V] + V_x.$$
It remains to note that the matrices $\ [P^{-1}P_t, U]\ $ and
$\ [P^{-1}P_x, V]\ $ are off-diagonal.
Hence, $$U_t=V_x.$$
The Lemma 1 is proved.
\vskip .1 in

Thus, besides three evident conservation laws with the densities
$a, b, c$ the system (1.3) has also three conservation laws with
the densities
 $\ u^1, u^2, u^3, \ $  which are roots of the
characteristic equation
$$\det (\lambda E - A) = \lambda^3 - a\lambda^2-2b\lambda-c =0.$$
By virtue of the obvious linear relation $a=u^1+u^2+u^3$ among them
only five conservation laws with the densities $\ u^1, u^2, u^3, b, c\ $
are linearly independent.
It is easy to show that the system (1.3)
has no other conservation laws of hydrodynamic type,
i.e., with densities of the form $\ h(a,b,c).$

Let us go over in the equations (1.3) from the variables
 $\ a, b, c\ $
to new field variables $\ u^1, u^2, u^3,\ $
connected with $\ a, b, c \ $ by the Vi\`ete formulas
$$a = u^1 +u^2+u^3,\ \ b =-{1 \over 2} (u^1 u^2 + u^2 u^3 + u^3 u^1),\
\ c = u^1 u^2 u^3.$$
For shortening of calculations we note that
the matrices $A$ and $B$ are connected by the relation
$$B=A^2-aA-bE.$$
Hence, the same relation is valid for the corresponding
diagonal matrices $U$ and $V$ :
$$V=U^2-aU-bE.$$
Substituting the expressions for $a$ and $b$
and using the equation (2.2), we obtain the following
representation for the system (1.3)
$$U_t = (U^2 - (u^1 +u^2 +u^3)U + {1 \over 2} (u^1 u^2 + u^2 u^3 +
u^3 u^1) E )_x$$
or, in components,
$${\pmatrix {u^1 \cr u^2 \cr u^3 \cr }}_t =
{1 \over 2} {\pmatrix {u^2 u^3  -u^1 u^2  -u^1 u^3 \cr
u^1 u^3  -u^2 u^1 -u^2 u^3 \cr u^1 u^2 -u^3 u^1 - u^3 u^2 \cr}}_x =
{1 \over 2} {\pmatrix {1&-1&-1 \cr -1& 1& -1 \cr -1& -1& 1 \cr}}
{d \over dx} {\pmatrix {{{\partial h} / {\partial u^1}} \cr
{{\partial h} / {\partial u^2}} \cr {{\partial h} / {\partial u^3}} \cr}},
  \eqno{(2.3)}$$
where $\ h=c=u^1 u^2 u^3.\ $  Hence,
the system under consideration is Hamiltonian
with the Hamiltonian operator
$$M={1 \over 2} {\pmatrix {1&-1&-1\cr -1&1&-1 \cr -1&-1&1 \cr}}
{d \over dx} \eqno{(2.4)}$$
and the Hamiltonian functional $\ H= \int cdx = \int u^1 u^2 u^3 dx.$
Density of momentum and annihilators of the corresponding Poisson
bracket has the following form:

$2b = -u^1 u^2 - u^2 u^3 - u^3 u^1 \ \ \ $ (density of momentum),

$u^1, u^2, u^3\ \ \ \ $ (annihilators).

In the initial variables $\ a, b, c\ $ the Hamiltonian operator (2.4)
is expressed as

$$M= {\pmatrix {-{3 \over 2} & {1 \over 2} a & b \cr
\noalign{\vskip 7pt}
{1 \over 2} a &b & {3 \over 2} c \cr
\noalign{\vskip 7pt}
b & {3 \over 2} c & 2(b^2 -ac) \cr}}{d \over dx} +
{\pmatrix {0 & {1 \over 2} a_x & b_x \cr
\noalign{\vskip 7pt}
0 & {1 \over 2} b_x & c_x \cr
\noalign{\vskip 7pt}
0 & {1 \over 2} c_x & (b^2 -ac)_x \cr}}.$$

We recall that local nondegenerate
Hamiltonian operators of hydrodynamic type
were introduced and studied by Dubrovin and Novikov (see [5]).
It was proved that
an operator of the form
$$P^{ij} = g^{ij} (u) {d \over dx} + b^{ij}_k (u) u^k_x,
\ \ \ \det [g^{ij}(u)] \neq 0,$$
is Hamiltonian if and only if

\item{(1)} $g^{ij} (u)$ is a metric of zero Riemannian curvature
(i.e., simply a flat metric);

\item{(2)} $b^{ij}_k (u) = - g^{is}(u) \Gamma ^j_{sk} (u)$,
where $\Gamma ^j_{sk} (u)$ are the coefficients of the
differential geometric connection generated by the metric
$g^{ij}$, i.e., the only symmetric connection compatible with
the metric (the Levi-Civita connection).

Correspondingly, Hamiltonian systems of hydrodynamic type,
which were considered by Dubrovin and Novikov [5],
have the form
$$u^i_t = P^{ij} {\partial H \over \partial u^j},$$
where $H= \int h(u)dx$ is a functional of hydrodynamic type.
Nonlocal generalizations of the Hamiltonian theory of
systems of hydrodynamic type were discovered in [13] (see also
[14-15]). Efficient theory of integrability
of diagonalizable
Hamiltonian systems of hydrodynamic type, i.e., in other words,
Hamiltonian systems of hydrodynamic type, which can be
reduced to Riemann invariants:
$$R^i_t =v^i(R) R^i_x,$$
was built by Tsarev [4].
All such systems have infinite
number of conservation laws
and commuting flows of hydrodynamic type and
can be integrated by generalized hodograph method.
However, it was shown in the papers [2,3] that
the system (1.3) do not possess Riemann invariants.
This explains, in particular, the fact that the system
(1.3) has only finite number of hydrodynamic type integrals.

General theory of integrability of nondiagonalizable (i.e.,
not possessing Riemann invariants) Hamiltonian systems of
hydrodynamic type started to develop in [6-9].
For three-component systems there were obtained final results.

\vskip .1 in
 {\bf Theorem 1 }[7,8].{\it Nondiagonalizable
Hamiltonian (with nondegenerate Poisson bracket of hydrodynamic type)
$3 \times 3$ system of hydrodynamic type
is integrable if and only if it is weakly nonlinear.}
\vskip .1 in

We recall that a system of hydrodynamic type
$$u^i_t = v^i_j(u) u^j_x, \ \ i,j=1,...,n, \eqno{(2.5)} $$
is called weakly nonlinear if for eigenvalues
 $\lambda ^i (u)$ of matrix $v^i_j (u)$
 the following relations
$$L_{\vec {X}^i} (\lambda ^i) =0,$$
where $L_{\vec {X}^i}$ is the Lie derivative along eigenvector
 $\vec {X}^i$ corresponding to eigenvalue
$\lambda ^i$, are satisfied for any $\ i = 1,...,n$.

There exists a simple and efficient criterion of weak nonlinearity,
which does not appeal to eigenvalues and eigenvectors.
\vskip .1 in
{\bf Proposition} [7]. {\it A system of hydrodynamic type (2.5)
is weakly nonlinear if and only if
$$(\grad f_1) v^{n-1} + (\grad f_2) v^{n-2} + ... + (\grad f_n)E = 0,$$
where $f_i$ are coefficients of characteristic polynomial
$$\det (\lambda \delta ^i_j - v^i_j (u)) = \lambda ^n + f_1 (u)
\lambda ^{n-1} + f_2 (u) \lambda ^{n-2} + ... + f_n (u), $$
and $v^n$ denotes $n$-th power of matrix $v^i_j$.}

As it was shown in [2,3], the systems (1.3) and (1.4) are weakly
nonlinear.

\vskip .3 in
\centerline{\bf \S \ 3  Integrable Hamiltonian $3 \times 3\ $
systems of hydrodynamic type,}
\centerline{\bf which do not possess
Riemann invariants}

\vskip .2 in

Consider a system of hydrodynamic type
$$u^i_t = v^i_j(u) u^j_x.  \eqno{(3.1)}$$
Let $\lambda ^i (u)$ be eigenvalues of matrix $v^i_j,$
i.e., roots of characteristic equation $\det (v^i_j (u) - \lambda
\delta ^i_j) =0$
(we assume that the system under consideration is
strictly hyperbolic, i.e., all the roots of the characteristic
equation are real and distinct).
Denote by $\vec {l}^{\, \, i} (u) = (l^i_1,...,l^i_n)$
the left eigenvector of the matrix $v^i_j,$ which corresponds to
eigenvalue $\lambda ^i,$ i.e., $l^i_k v^k_j = \lambda ^i l^i_j.$
Introduce 1-forms $\omega ^i = l^i_k du^k \ (i=1,...,n).$
We emphasize that 1-forms $\omega ^i$
are defined up to normalization
 $\omega ^i \mapsto  p^i \omega ^i, \ p^i \neq 0.$
It is easy to verify that the equations (3.1) can be rewritten
in the form of system of exterior equations
$$\omega ^i \wedge (dx + \lambda ^i dt) =0,\ \ \ i=1,...,n. \eqno{(3.2)}$$

For the being investigated system (2.3) eigenvalues $\lambda ^i$
and corresponding them left eigenvectors $\vec l^{\, \, i}$
have the form:
$$\lambda ^1 = -u^1,\ \ \ \vec {l}^{\, \, 1} = (u^2-u^3, u^1-u^3, u^2-u^1),$$
$$\lambda ^2 = -u^2,\ \ \ \vec {l}^{\, \, 2} = (u^2-u^3, u^1-u^3, u^1-u^2),$$
$$\lambda ^3 = -u^3,\ \ \ \vec {l}^{\, \, 3} = (u^2-u^3, u^3-u^1, u^2-u^1).$$
Hence, the equations (2.3) can be expressed as
$$\omega ^i  \wedge (dx -u^i dt) =0,\ \ i=1,2,3,$$
where
$$\omega ^1 = (u^2-u^3)du^1 + (u^1-u^3)du^2 + (u^2-u^1)du^3,$$
$$\omega ^2 = (u^2 -u^3)du^1 + (u^1 -u^3)du^2 + (u^1-u^2)du^3,$$
$$\omega ^3 = (u^2-u^3)du^1 + (u^3-u^1)du^2 + (u^2-u^1)du^3. \eqno{(3.3)}$$

Let $B(u)dx + A(u)dt$ and $N(u)dx+M(u)dt$ be
two hydrodynamic type integrals of a system of hydrodynamic type
(3.1), i.e.,
the differential 1-forms are closed
along solutions of the system.
Go over from the variables $x, t$
to new independent variables
$\tilde {x}, \tilde {t}$ by means of the following formulas
$$d \tilde {x}=Bdx+Adt,$$
$$d \tilde {t} =Ndx+Mdt.  \eqno{(3.4)}$$
Then the system (3.1) transforms to the form
$$u^i_{\tilde {t}} ={\tilde {v}}^i_j(u) u^j_{\tilde {x}},$$
where the matrix $\tilde {v}$ is related with $v$ by the formula
$$\tilde {v} = (Bv-AE)(ME-Nv)^{-1}.$$
Using the language of exterior equations we can rewrite
the transformed system in the following form
$$\omega ^i \wedge (d \tilde {x} +{\tilde {\lambda}} ^i d \tilde {t}) =0,
\ \ \ i =1,...,n,
 \eqno{(3.5)}$$
where
$${\tilde {\lambda}} ^i = {{\lambda ^i B - A} \over {M - \lambda ^i N}}.
\eqno{(3.6)}$$
Hence, the 1-forms
$\omega ^i$
do not change after the transformations of the form (3.4)
while the eigenvalues
$\lambda ^i$
transform in according with the formula (3.6).
\vskip .1 in
{\bf Theorem 2} [7,8].
{\it If a $3 \times 3$ system of hydrodynamic type (3.1)
is weakly nonlinear and Hamiltonian (with nondegenerate
Poisson bracket of hydrodynamic type) then
there exists a pair of integrals (3.4) of this system such that
the corresponding transformed system has constant eigenvalues
${\tilde {\lambda}}^i$,
which can be considered equal to
\ 1, -1, 0 \  without loss of generality.}
\vskip .1 in

For the system (2.3) the transformation (3.4), the existence  of which
is established by Theorem 2, has the following form:
$$d \tilde {x} = Bdx + Adt = (u^1 -u^2)dx +u^3(u^2-u^1)dt,$$
$$d \tilde {t} = Ndx + Mdt = (2u^3 -u^1-u^2)dx +(2u^1 u^2 -u^1 u^3
-u^2 u^3)dt.  \eqno{(3.7)}$$

According to the formula (3.6) the transformed eigenvalues will be
equal to
1, -1, 0, correspondingly.
Hence, in new independent variables
 $\tilde {x}, \tilde {t}$
the system (2.3) can be rewritten in the form
$$\omega ^1 \wedge (d \tilde {x} + d \tilde {t} ) =0,\ \
\omega ^2 \wedge (d \tilde {x} -d \tilde {t}) =0,\ \
\omega ^3 \wedge d \tilde {x} =0.$$
\vskip .1 in

{\bf Theorem 3} [7,8].
{\it If a $3 \times 3$ system of hydrodynamic type (3.1)
is nondiagonalizable, weakly nonlinear and Hamiltonian
(with nondegenerate Poisson bracket of hydrodynamic type)
then the corresponding 1-forms
$\omega ^1, \omega ^2, \omega ^3$
can be normalized such that they will satisfy either
the structural equations of the group $SO(3)$:
$$d \omega ^1 = \omega ^2 \wedge \omega ^3, \ \
d \omega ^2 = \omega ^3 \wedge \omega ^1, \ \
d \omega ^3 = \omega ^1 \wedge \omega ^2,  \eqno{(3.8)}$$
if the signature of the metric, determining the Poisson bracket
of hydrodynamic type,
is Euclidean, or the structural equations of the group $SO(2,1)$:
$$d \omega ^1 = \omega ^2 \wedge \omega ^3, \ \
d \omega ^2 = \omega ^3 \wedge \omega ^1, \ \
d \omega ^3 = - \omega ^1 \wedge \omega ^2  \eqno{(3.9)}$$
(in the case of Lorentzian signature of the metric).}

\vskip .1in
For the system (2.3) the signature of the metric
of the Poisson bracket (2.4) is Lorentzian.
Hence, the forms (3.3) can be normalized such that
they will satisfy
the structural equations (3.9).
Desired normalization has the form ( we will not introduce
a new notation for the normalized 1-forms $\omega ^i$):
$$\omega ^1 ={{(u^2 - u^3)du^1 + (u^1-u^3)du^2 + (u^2-u^1)du^3}
\over {2(u^2-u^3) \sqrt {(u^2-u^1)(u^3-u^1)}}},$$
$$\omega ^2 = {{(u^2-u^3)du^1 + (u^1-u^3)du^2 +(u^1-u^2)du^3}
\over {2(u^3-u^1) \sqrt {(u^2-u^1)(u^2-u^3)}}},$$
$$\omega ^3 = {{(u^2-u^3)du^1 + (u^3-u^1)du^2 +(u^2-u^1)du^3}
\over {2(u^2-u^1) \sqrt {(u^3 -u^1)(u^2-u^3)}}}  \eqno{(3.10)}$$
(for definiteness we here consider $u^1<u^3<u^2$).

One can verify by direct check that the 1-forms (3.10)
satisfy to the structural equations (3.9).

Thus, according to theorems 2 and 3 any nondiagonalizable
weakly nonlinear Hamiltonian (with nondegenerate
Poisson bracket of hydrodynamic type)
$3 \times 3$ system of hydrodynamic type can be
reduced to the canonical form
$$\omega ^1 \wedge (d \tilde {x} + d \tilde {t} ) =0,
\ \ \omega ^2 \wedge (d \tilde {x} -d  \tilde {t}) =0,\ \
\omega ^3 \wedge d \tilde {x} =0  \eqno{(3.11)}$$
by suitable transformation (3.4).
Moreover, one can consider that for the forms $\omega ^i$
the structural equations (3.8) or (3.9) are satisfied
(note that the transformations of the form (3.4) do not change
the structural equations). Introducing in the equations (3.11)
the variables $p^1, p^2, p^3$ in according to the formulas
(see [6,7])
$$\omega ^1 = p^1 (d \tilde {x} + d \tilde {t}), \ \
\omega ^2 =p^2 (d \tilde {x} - d \tilde {t}),\ \
 \omega ^3 = p^3 d \tilde {x}. \eqno{(3.12)}$$
and substituting (3.12) in the structural equations (for
definiteness in (3.9)) we obtain the integrable system of
three waves
$$p^1_{\tilde {t}} - p^1_{\tilde {x}} = - p^2 p^3,$$
$$p^2_{\tilde {t}} + p^2_{\tilde {x}} = - p^1 p^3,$$
$$p^3_{\tilde {t}} = - 2 p^1 p^2.  \eqno{(3.13)}$$
\vskip .1 in
{\bf Remark}. If use explicit coordinate representation
of the 1-forms $\omega ^i = l^i_k(u) d u^k,$ then for $p^i$
we obtain expressions of the form $p^i = l^i_k (u) u^k_{\tilde {x}}$.
Hence, the change of variables from $u^i$ to $p^i$ is
a differential substitution of the first order.
\vskip .1 in

Summing up the described construction for the considered system
we can present the transition from the equations (2.3) to
the system of three waves (3.13) as two the following steps:

1. The change of variables from $x, t$
to new independent variables $\tilde {x},
\tilde {t}$ in according to the formula (3.7).

2. The change of the field variables from $u^1, u^2, u^3$ to $p^1, p^2, p^3$
in according to the following formulas (compare with (3.10)):

$$p^1 = {{(u^2-u^3) u^1_{\tilde {x}} + (u^1 - u^3) u^2_{\tilde {x}}
+ (u^2 -u^1) u^3_{\tilde {x}}} \over {2(u^2-u^3) \sqrt {(u^2-u^1)
(u^3  -u^1)}}},$$
$$p^2 = {{(u^2-u^3) u^1_{\tilde {x}} + (u^1 - u^3) u^2_{\tilde {x}}
+ (u^1 -u^2) u^3_{\tilde {x}}} \over {2(u^3-u^1) \sqrt {(u^2-u^1)
(u^2  -u^3)}}},$$
$$p^3 = {{(u^2-u^3) u^1_{\tilde {x}} + (u^3 - u^1) u^2_{\tilde {x}}
+ (u^2 -u^1) u^3_{\tilde {x}}} \over {2(u^2-u^1) \sqrt {(u^3-u^1)
(u^2  -u^3)}}}. \eqno{(3.14)}$$

Thus, any solution of the integrable three wave system (3.13)
generates solutions of the equations of associativity (1.1).
\vskip .3 in

\centerline{\bf \S \ 4.  Relation between the systems (1.3) and (1.4)}

\vskip .2 in

The spectral problem corresponding to the system (1.4)
has the form
$$\Psi _x = z A\Psi = z {\pmatrix {0&1&0 \cr
0&b&a \cr 1&c&b \cr}} \Psi,$$
$$\Psi _t = z B\Psi = z {\pmatrix {0&0&1 \cr
1&c&b \cr
0& {(1+bc)/a} &c}} \Psi.  \eqno{(4.1)}$$
It is easy to verify that the matrix $B$ is related with $A$
by the formula
$$B={1 \over a} (A^2 - bA).$$
Compatibility condition for the spectral problem (4.1)
$$A_t=B_x,$$
rewritten in terms of eigenvalues of matrices
$A$ and $B$ (see Lemma 1, \S \ 2), has the form:
$$w^i_t = \biggl ( {1 \over a} ((w^i)^2 - w^i b) \biggr ) _x,
\eqno{(4.2)}$$
where $w^i$ are eigenvalues of the matrix $A$, i.e.,
the roots of the characteristic equation
$$\det (\lambda E-A) = \lambda ^3 -2b\lambda ^2 +(b^2-ac)\lambda -a=0.$$
Expressing $a$ and $b$ by the Vi\`ete formula
$$b = {1 \over 2} (w^1 +w^2 +w^3),\ \ \ a = w^1 w^2 w^3$$
and substituting the expressions in (4.2) we obtain
the explicit representation of the equations (1.4) in
the coordinates $w^i$:
$${\pmatrix {w^1 \cr w^2 \cr w^3 \cr}}_t = {1 \over 2}
{\pmatrix {(w^1 - w^2 - w^3) / w^2 w^3 \cr
(w^2 -w^1 -w^3) / w^1 w^3 \cr (w^3-w^1-w^2)/w^1 w^2 \cr}}_x.
 \eqno{(4.3)}$$

Note that the integrable systems of hydrodynamic type
(1.4) and (4.3) do not possess local Hamiltonian structures
of hydrodynamic type (the Poisson brackets of Dubrovin-Novikov
type [5]). Corresponding them Hamiltonian structures of hydrodynamic
type are strictly nonlocal ([14--16]).

We show the explicit relation between the systems (2.3) and (4.3).
For this we shall go over in the equations (2.3) from $x,\ t$
to new independent variables $\tilde {x},\ \tilde {t}$ in according to
formulas
$$d \tilde {x} = - {1 \over 2} (u^1 u^2 + u^1 u^3 + u^2 u^3) dx +
u^1 u^2 u^3 dt,\ \ \ d \tilde {t} = dx. \eqno{(4.4)}$$
After the transformation (4.4) the system (2.3) has the form
$${\pmatrix {{1 / {u^1}} \cr {1 / {u^2}} \cr {1 / {u^3}} \cr
}} _{\tilde t} = {1 \over 2} {\pmatrix { {(u^2
 u^3) / u^1} - u^2 -u^3 \cr {(u^1 u^3)/ u^2} -u^1 - u^3 \cr
{(u^1 u^2)/ u^3} - u^1 -u^2}} _{\tilde x}, $$
i.e., as it is easy to see, it coincides with (4.3) after
the transformation
$$w^i = {1 \over u^i}. \eqno{(4.5)} $$

Using the language of the initial equations (1.1) and (1.2)
we can present
the transformations (4.4) and (4.5)
in the following way:
the equation
$$f_{ttt} = f^2_{xxt} -f_{xxx}f_{xtt}$$
goes over into the equation
$$
{\tilde f}_{{\tilde {x}} {\tilde {x}} {\tilde {x}}}
{\tilde f}_{{\tilde {t}}\, {\tilde {t}}\, {\tilde {t}}}-
{\tilde f}_{{\tilde {x}} {\tilde {x}} {\tilde {t}}}
{\tilde f}_{{\tilde {x}} {\tilde {t}}\, {\tilde {t}}}
=1$$
after the transformation
$$\tilde {x} = f_{xt}, \ \ \ \tilde t =x,$$
$$
{\tilde {f}} _{{\tilde {x}} {\tilde {x}}}=t,\ \
{\tilde {f}} _{{\tilde {x}} {\tilde {t}}}=-f_{xx},\ \
{\tilde {f}} _{{\tilde {t}}\, {\tilde {t}}}=f_{tt}.\ \ \eqno{(4.6)} $$
Note that this transformation, connecting solutions of the equations
associativity (1.1) and (1.2),
is not contact.

\vskip .3 in
\centerline{\bf References}

\vskip .2 in
\noindent
\item{[1]} Dubrovin B., Geometry of 2D topological field theories.
Preprint SISSA--89/94/FM, SISSA, Trieste (1994).
\item{[2]} Mokhov O.I., Differential equations of associativity
in 2D topological field theories and geometry of nondiagonalizable
systems of hydrodynamic type. In: Abstracts of Internat. Conference on
Integrable Systems "Nonlinearity and Integrability: from Mathematics
to Physics", February 21--24, 1995, Montpellier, France (1995).
\item{[3]} Mokhov O.I., Symplectic and Poisson geometry on loop
spaces of manifolds and nonlinear equations. Advances in Soviet Mathematics.
S.P.Novikov seminar. Ed. S.P.Novikov. {\bf 22}, AMS, Providence, USA
(1995).
\item{[4]} Tsarev S.P., Geometry of Hamiltonian systems of
hydrodynamic type. The generalized hodograph method. Izvestiya Akad. Nauk
SSSR, Ser. mat., {\bf 54}, No. 5, 1048--1068 (1990).
\item{[5]} Dubrovin B.A. and Novikov S.P.,  Hydrodynamics of
weakly deformed soliton lattices. Differential geometry and
Hamiltonian theory. Uspekhi Mat. Nauk, {\bf 44}, No. 6, 29--98 (1989).
\item{[6]} Ferapontov E.V., On integrability of $3 \times 3$
semi-Hamiltonian hydrodynamic type systems which do not possess
Riemann invariants. Physica D, {\bf 63}, 50--70 (1993).
\item{[7]} Ferapontov E.V., On the matrix Hopf equation and
integrable Hamiltonian systems of hydrodynamic type, which do not
possess Riemann invariants. Phys. Lett. A, {\bf 179}, 391--397 (1993).
\item{[8]} Ferapontov E.V., Dupin hypersurfaces and integrable
Hamiltonian systems of hydrodynamic type which do not possess
Riemann invariants. Diff. Geometry and Appl., (1995).
\item{[9]} Ferapontov E.V., Several conjectures and results
in the theory of integrable Hamiltonian systems of hydrodynamic type,
which do not possess Riemann invariants. Teor. and Mat. Physics,
{\bf 99}, No. 2, 257--262 (1994).
\item{[10]}Witten E., On the structure of the topological
phase of two-dimensional gravity. Nucl. Physics B,
{\bf 340}, 281--332 (1990).
\item{[11]} Dijkgraaf R., Verlinde H. and Verlinde E.,
Topological strings in $d < 1$.
Nucl. Physics B, {\bf 352}, 59--86 (1991).
\item{[12]} Witten E., Two-dimensional gravity and intersection
theory on moduli space. Surveys in Diff. Geometry, {\bf 1},
 243--310 (1991).
\item{[13]} Dubrovin B., Integrable systems in topological
field theory. Nucl. Physics B, {\bf 379}, 627--689 (1992).
\item{[14]} Mokhov O.I. and Ferapontov E.V., On the nonlocal
Hamiltonian hydrodynamic type operators connected with constant
curvature metrics. Uspekhi Mat. Nauk, {\bf 45}, No. 3, 191--192 (1990).
\item{[15]} Ferapontov E.V., Differential geometry of nonlocal
Hamiltonian operators of hydrodynamic type. Funkts. Analiz i ego
Prilozh., {\bf 25}, No. 3, 37--49 (1991).
\item{[16]} Mokhov O.I., Hamiltonian systems of hydrodynamic type and
constant curvature metrics. Phys. Letters A, {\bf 166},
No. 3--4, 215--216 (1992).
\vskip .3 in

\bye